\documentclass[aps,prb,twocolumn,superscriptaddress,longbibliography]{revtex4-1}
\usepackage[colorlinks=true,citecolor=blue,linkcolor=blue,breaklinks=true]{hyperref}

\usepackage{color,graphicx}
\usepackage{bm}
\usepackage{amsmath}
\usepackage{amssymb}
\usepackage{times}

\allowdisplaybreaks

\begin{document}
	
\newcommand {\up} {\ensuremath{\uparrow}}
\newcommand {\dn} {\ensuremath{\downarrow}}
\newcommand {\no} {\nonumber}
\newcommand{\rcol} {\textcolor{red}}
\newcommand{\bcol} {\textcolor{blue}}
\newcommand{\lt} {\left}
\newcommand{\rt} {\right}
\newcommand{\nbr} {\ensuremath{\langle ij \rangle}}
\newcommand{\bk} {\ensuremath{{\rm \bf k}}}
\newcommand{\bq} {\ensuremath{{\rm \bf q}}}

%
\title{Phonon-induced modification of quantum criticality}
\author{Abhisek Samanta}\email{abhiseks@campus.technion.ac.il}
\affiliation{ Physics Department, Technion, Haifa 32000, Israel}
\author{Efrat Shimshoni}\email{Efrat.Shimshoni@biu.ac.il}
\affiliation{Department of Physics, Bar-Ilan University, Ramat Gan 52900, Israel}
\author{Daniel Podolsky}\email{podolsky@physics.technion.ac.il}
\affiliation{ Physics Department, Technion, Haifa 32000, Israel}

\date{\today }

\begin{abstract}
We study the effect of acoustic phonons on the quantum phase transition in the O($N$) model. We develop a renormalization group analysis near (3+1) space-time dimensions and derive the RG equations using an $\epsilon$-expansion. Our results indicate that when the number of flavors of the underlying O($N$) model exceeds a critical number $N_c=4$, the quantum transition remains second-order of the Wilson-Fisher type while, for $N\le 4$, it is a weakly first-order transition.  We characterize this weakly first-order transition by a length-scale $\xi^*$, below which the behavior appears to be critical.  At finite temperatures for $N\le 4$, a tricritical point separates the weakly first-order and second-order transitions.
\end{abstract}
\pacs{XXXX}

\maketitle

\section{Introduction}
The fate of the second-order quantum phase transition in the presence of lattice vibrations is an intriguing question which still remains to be completely understood.  The transition between ordered and disordered phases in magnets, superfluids, charge density waves, {\em etc.} are typically studied within lattice models which assume the lattice to be static, both in the classical and quantum cases~\cite{sachdev}. However, acoustic phonons are ubiquitous in realistic solid state systems, and their gaplessness gives reason to expect fundamental changes of the  standard critical behavior.

The effect of phonons in classical phase transitions has been studied extensively and it was a topic of controversy for many years~\cite{rice,domb,fisher}. Using a simplified continuum model for the elastic lattice, Larkin and Pikin derived a criterion by which the second-order transition becomes first-order whenever the magnetic specific heat becomes large~\cite{larkin_pikin}.  Intuitively, this results from the tendency of the system to gain energy by making distortions in the lattice.  The Larkin-Pikin criterion has been used extensively in the literature in the study of different models~\cite{bergman_halperin,imry_aharony,bruno_sak}. This picture was revisited by Aharony \cite{aharony}, who showed through a renormalization group (RG) analysis in $d=4-\epsilon$ space dimensions that, contrary to the Larkin-Pikin criterion, the transition may remain second-order provided the phonon coupling is weak enough.  

More recently, the focus has shifted to understanding the role of phonons in quantum phase transitions~\cite{meyer,eberg,alberton,pchandra,moon}.  This is motivated by experiments on new platforms, such as interacting atoms, ions, and dipoles in a trap, which open possibilities to study quantum phase transitions in systems with soft lattices.  On the solid state front,  experiments on ferroelectric materials ~\cite{rowley,nova,ahadi,aseginolaza,brando} further motivate this study.  We note that a prominent effect of the coupling to phonons is the explicit breaking of Lorentz invariance, which is often present in the effective field theory of the quantum  O($N$) model.  Lorentz violating terms of certain types have been shown to alter the critical behavior~\cite{vieira}.

A number of theoretical analyses have looked at the quantum O($N$) model coupled to phonons in $D=1+1$ space-time dimensions.  An RG analysis was performed on a quantum wire~\cite{meyer}, where it was shown that the transition could be second-order or first-order depending on the ratio of the spin-wave and phonon velocities. This analysis was extended in Ref.~\onlinecite{alberton} and supplemented by a numerical verification using DMRG calculations~\cite{alberton}.  Under specific conditions, the 1+1 dimensional problem has been shown to support emergent supersymmetric quantum criticality~\cite{eberg,moon}. However, understanding the effect of phonons on quantum critically in higher dimensions remains a challenge.

Recently, the Larkin-Pikin criterion for magnetic transitions was generalized by including the quantum fluctuations~\cite{pchandra}. There, it was shown that the universality of the Wilson-Fisher fixed point remains robust in $D=3+1$, where the spin-phonon coupling term is argued to be marginally irrelevant. Below three space dimensions, however, such benign behavior of the coupling to phonons is no longer guaranteed. 

\begin{figure}
\includegraphics[width=0.98\columnwidth]{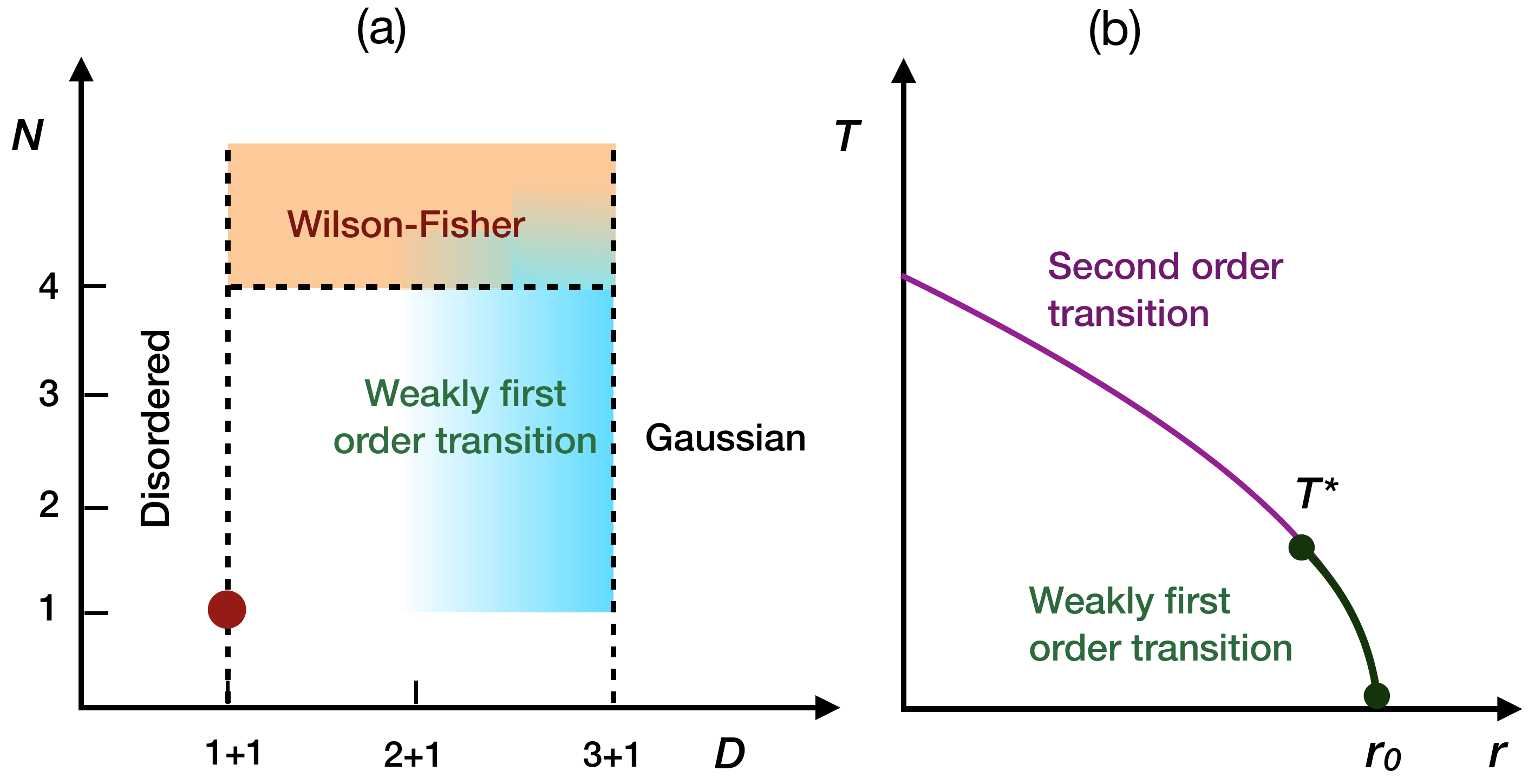}
\caption{(a): The nature of the quantum phase transition of the O($N$) model in presence of acoustic phonons is shown in the $N-D$ plane, where $N$ is the number of flavors of the O($N$) spins and $D$ is the space-time dimensions. We find a second-order transition (governed by Wilson-Fisher fixed point) for $N>4$. The transition turns weakly first-order for $N\le4$. The quantum phase transition in $D=1+1$ has been studied previously~\cite{meyer,alberton} (shown by a red point in the phase diagram). (b): We characterize the weakly first-order transition for $1<N\leq4$ by a lengthscale $\xi^*$, which can be related to a temperature scale $T^*$, below which the transition becomes first-order.}
\label{Fig:phasediag}
\end{figure}

In the present work, we show that phonons can indeed have a strong effects on quantum criticality below $D=3+1$ dimensions. To this end we study the quantum O($N$) model weakly coupled to acoustic phonons by performing an RG analysis near (3+1) dimensions, utilizing an $\epsilon$-expansion. Integrating out the phonons in our model leads to a non-local interaction in the effective action for the O($N$) order parameter field, which we analyze in detail. Our main finding is the presence of a critical number of flavors, $N_c=4$: when the number of flavors $N$ exceeds this value, the transition remains second-order governed by the standard Wilson-Fisher fixed point, while it turns weakly first-order below this critical number [Fig.~\ref{Fig:phasediag}(a)]. We characterize this weakly first-order transition by a length-scale $\xi^*$, which diverges exponentially as $\epsilon$ approaches zero or the coupling to phonons becomes progressively smaller. This length-scale can be heuristically related to the temperature scale $T^*$ of a tricritical point, below which the transition turns from second-order to weakly first-order [Fig.~\ref{Fig:phasediag}(b)].  For the Ising model, $N=1$, the divergent specific heat of the rigid-lattice model has been shown to lead to a thermal transition that is at least weekly first order\cite{bergman_halperin}.  Hence, the tricritical point in Fig.~\ref{Fig:phasediag}(b) is predicted to occur for $1<N\le 4$.

The rest of the paper is organized as follows: In Section II we introduce the coupled spin-phonon model, and derive an effective action for the spins resulting from integration over the phonons. In Section III we present the RG analysis of the effective action in $D=3+1-\epsilon$ dimensions, and derive the RG equations. In Section IV we show the solutions of the RG equations and discuss the results. Finally, in Section V we include a brief overview of our key results and concluding remarks.
\section{Coupled spin-phonon model}
We consider a soft O($N$) quantum spin model in $d$ space dimensions, with Euclidean action
\begin{align}
{\mathcal S} &= \int\!\! d\tau \lt[\sum_i \lt\{(\partial_\tau {\vec \phi_i})^2+r \vec\phi_i^2 +\frac{U_0}{N}\lt({\vec \phi}_i^2\rt)^2\rt\}\rt.\no\\
&\quad\quad \lt.-\sum_{\langle ij\rangle} J({\rm \bf R}_{ij}){\vec\phi_i}\cdot{\vec\phi_j}\rt]
\label{eq:SFirst}
\end{align}
where $i$ and $j$ are sites of the lattice, and we use arrows to indicate vectors in the internal O($N$) space and boldface letters to indicate vectors in real space.  Here,
$J$ is the magnetic exchange between spins, which we assume to be dependent on the separation ${\rm \bf R}_{ij}={\rm \bf R}_i-{\rm \bf R}_j$ between nearest neighbors $\langle i j\rangle$.  We also assume relativistic invariance (no first-order time derivatives) in the spin sector, as would be present {\em e.g.} in Heisenberg ferromagnets or in non-particle-hole symmetric superfluids.  For instance, Eq. (\ref{eq:SFirst}) can describe Heisenberg antiferromagnets on bipartite lattices, with the field $\vec \phi$ representing the N\'eel vector.


We now introduce phonons by allowing the lattice to be dynamical.  Then, the position at site $i$ can be written in terms of the displacement ${\bf u}_i$ from the equilibrium position ${\bf R}_i^{(0)}$ as ${\bf R}_i={\bf R}_i^{(0)}+{\bf u}_i$.  This gives rise to a quantum O($N$) model
coupled to gapless phonons, in the form of the Wagner-Swift
Hamiltonian~\cite{aharony}. The corresponding Euclidean action is given by
\begin{align}
\label{eq:S}
{\mathcal S} &={\mathcal S}_s  + {\mathcal S}_p + {\mathcal S}_{sp} \\
\label{eq:Ss}
{\mathcal S}_s &\!=\!\! \int\!\! d\tau \!\!\lt[ \sum_i\!\!\lt\{(\partial_\tau {\vec \phi_i})^2 
\!+\! r\vec\phi_i^2 \!+\!\frac{U_0}{N} \lt({\vec \phi}_i^2\rt)^2\rt\}\!-\!J_0 \sum_{\langle ij\rangle}\! {\vec\phi_i}\cdot{\vec\phi_j}\!\rt] \\ 
\label{eq:Sp}
{\mathcal S}_p &= \int\!\! d\tau \lt[ \frac{M}{2}\sum_i\lt(\partial_\tau {\bf u}_i\rt)^2
+ \frac{1}{2}\sum_{\langle ij\rangle}\kappa^{ab}_{ij}\ u^a_i u^b_j \rt] \\
\label{eq:Ssp}
{\mathcal S}_{sp} &= \int\!\! d\tau \lt[\sum_i\Psi^a_i u^a_i \rt]
\end{align}
where $J_0=J({\rm \bf R}_{ij}^{(0)})$ is independent of the bond $\langle ij \rangle$. In what follows, we adopt units where $J_0=1$.
$\mathcal S_p$ is the action of the harmonic phonons [Eq.~(\ref{eq:Sp})], where $M$ is the mass of the atoms and $\kappa^{ab}_{ij}=\kappa^{ab}({\rm \bf R}_{ij}^{(0)})$ is the elastic tensor of the lattice.  Here, we have expanded the spin-lattice interaction to first-order in the displacements ${\bf u}_i$.  Therefore,
\begin{eqnarray}
\Psi^a_i &=& -\sum_{\langle jl\rangle}\left.{\partial J({\rm \bf R}_{jl}) \over \partial R_i^a}\right|_{{\bf R}_{jl}^{(0)}} \ {\vec\phi_j}\cdot{\vec\phi_l}\; .  
\label{eq:Psiam}
\end{eqnarray}
A sum over repeated indices is implicit throughout the paper.

It is useful to write the action in Fourier space.  To do so, we assume $J({\rm \bf R}_{ij})=J(|{\rm \bf R}_{ij}|)$ and define
\begin{align}
g=\left. {\partial J({\bf R}_{ij}) \over \partial |{\bf R}_{ij}|}\right|_{{\bf R}_{ij}^{(0)}} ,    
\end{align}
which is independent of the bond ${\langle ij\rangle}$ and plays the role of the spin-phonon coupling constant.  Then, the Fourier transform of $\Psi_i^a(\tau)$ becomes
\begin{align}
\Psi^a({\bf q},\omega)\! &=\! -g \sum_{\langle ij\rangle} e^{i{\rm \bf q}\cdot{\rm \bf R}_i^{(0)}}\!\!
\lt(1- e^{-i{\rm \bf q}\cdot{{\bf R}_{ij}^{(0)}}} \rt)
{{ R}_{ij}^{a\,(0)} \over \lt|{\bf R}_{ij}^{(0)}\rt|}\ {\vec\phi_i}\cdot{\vec\phi_j} \no\\
&\!\!\!\!\!\!\!\!\!\!\!\!\!\!\!\!\!\!\!\!\!\!\! = {g\over 2}\!\int\!{d\omega'\over 2\pi} \!\sum_{{\bf k},\mu} \mu^a\! \lt(\!e^{-i{\rm \bf (q-k)}\cdot{\bf \mu}}\!\!-\!e^{i{\rm \bf k}\cdot{\mu}}\!\rt) {\vec\phi({\bf k},\omega')}\cdot{\vec\phi({\bf q}\!-\!{\bf k},\omega\!-\!\omega')} 
\end{align}
where $\vec\phi({\bf k},\omega)$ are the Fourier components of $\vec\phi_i(\tau)$, and $\mu$ is a shift by one lattice site in different directions.  Let us take an example of the square lattice, for which $\mu\in \{\pm \hat x, \pm \hat y\}$ in units where the lattice spacing is one. Then
\begin{align}
\Psi^a({\bf q},\omega) &=-g\!\int\!{d\omega'\over 2\pi}\sum_{\bf k} i\lt(\sin(q_a-k_a)+\sin k_a\rt) \no\\
&~~~~~~~~~~~~~~~~~~~~~~~~~~
\times{\vec\phi({\bf k},\omega')}\cdot{\vec\phi({\bf q}\!-\!{\bf k},\omega\!-\!\omega')} ,
\end{align}
which in the small $k$ and $q$ limit reduces to the following form,
\begin{align}
\Psi^a({\bf q},\omega) &= -ig\, q_a\!\int\!{d\omega'\over 2\pi} \sum_{\bf k} {\vec\phi({\bf k},\omega')}\cdot{\vec\phi({\bf q}\!-\!{\bf k},\omega\!-\!\omega')}\; .
\label{eq:psiq}
\end{align}

Since we are considering a quadratic theory for the phonons, we can integrate them out  to get 
an effective action for the spins.  This will come at the cost of making the effective spin interaction non-local.
The effective action for
the spins is ${\mathcal S}^{\rm eff}_{s} = {\mathcal S}_{s} + \delta{\mathcal S}_{s}$, where
\begin{widetext}
\begin{align}
{\mathcal S}_s &= \int\! {d\omega\over 2\pi} \sum_{\bf k} \lt(r+\omega^2+v^2\bk^2\rt) \vec\phi(\bk,\omega)\cdot\vec\phi(-\bk,-\omega)
+ {U_0\over N}\int{d\omega\over 2\pi} \sum_{\bf q}\lt|\int\! {d\omega'\over 2\pi} \sum_{\bf k} \vec\phi(\bk,\omega')\cdot\vec\phi({\bf q}-\bk,\omega-\omega')\rt|^2
\end{align}
and
\begin{align}
\delta{\mathcal S}_{s} &\!=\!\!\int\!\! {d\omega\over 2\pi}\sum_{\bf q}\!\sum_{ab} {\Psi^a({\bf q},\omega)\Psi^b(-{\bf q},-\omega)\over \lt[M\omega^2\delta^{ab}\!+\!\kappa^{ab}({\bf q})\rt]} 
\approx - \!\!\int\!\! {d\omega\over 2\pi}\sum_{\bf q}\! {g^2 {\bf q}^2 \over M(\omega^2+c^2{\bf q}^2)} \Big|\!\!\int\!\! {d\omega'\over 2\pi}\!\sum_{\bf k}  {\vec\phi({\bf k},\omega')}\!\cdot\!{\vec\phi({\bf q\!-\!k},\omega\!-\!\omega')}\Big|^2\,.
\end{align}
\end{widetext}
Here we assumed for simplicity an isotropic crystal with degenerate longitudinal and transverse phonons, for which $\kappa^{ab}({\bf q})=Mc^2{\bf q}^2\delta^{ab}$ where $c$ is the phonon velocity.
The partition function is therefore given by (up to an overall constant),
\begin{widetext}
\begin{align}
\label{eq:Z}
Z &= \int {\mathcal{D}}\phi\ e^{-S[\phi]}
~~= \int {\mathcal{D}}\phi \exp\lt[-{1\over 2} \int_{\omega,\bk}\lt(r+\omega^2+v^2\bk^2\rt) \vec\phi(\bk,\omega)\cdot\vec\phi(-\bk,-\omega) \rt. \no\\
&\lt. -{1\over N}\int_{\omega,\bq}\int_{\omega_1,\bk_1}\int_{\omega_2,\bk_2}\lt(\!U_0+W{c^2{\rm \bf q}^2\over \omega^2+c^2{\rm \bf q}^2}\!\rt)\!\phi^\alpha({\rm \bf k_1},\omega_1)\phi^\alpha({-{\rm \bf k_1}\!-\!{\rm \bf q}},-\omega_1\!-\!\omega)\phi^\beta({{\rm \bf k_2}\!+\!{\rm \bf q}},\omega_2\!+\!\omega)\phi^\beta(-{\rm \bf k_2},\!-\omega_2)\! \rt]
\end{align}
\end{widetext}
where $W=-{g^2N\over M}<0$, and we have used the short-hand notations $\int_{\omega,\bk}=\int^\infty_{-\infty} {d\omega\over 2\pi} \int_{|\bk|<\Lambda} {d^dk\over (2\pi)^d}$ and
\begin{eqnarray}
\mathcal{D}\phi&=&\prod_{\omega}\prod_{|\bk|<\Lambda}\prod_{\alpha=1}^N 
d\phi^\alpha(\bk,\omega)\,.
\end{eqnarray}
All momentum integrals are bounded by a cut-off $\Lambda$, set by the lattice spacing $a$, {\em i.e.} $\Lambda\sim {1\over a}$.
Note that due to the $W$ term, the action does not have Lorentz invariance.  Therefore, the cut-off only applies to the wave-vector, and we integrate over modes of all frequencies.

The $W$ term provides a correction to the standard $\phi^4$-model which is non-local in space-time. In momentum space, this is manifested by the dependence on transferred wave-vector and frequency of the effective interaction
\begin{eqnarray}
U_{\mathrm{eff}}(\bq,\omega)=U_0+W\frac{c^2\bq^2}{\omega^2+c^2\bq^2}\,.
\label{eq:Ueff}
\end{eqnarray}
Its naive scaling dimension at the Gaussian fixed point is the same as that of $U_0$, {\em i.e.} $D-4$ ($D=d+1$ being the space-time dimension). Hence, the upper critical space-time dimension is 4, which 
implies that we can perform a controlled expansion in $\epsilon=4-D$ (or, equivalenty, $\epsilon=3-d$). However, the non-analytical nature of $U_{\mathrm{eff}}(\bq,\omega)$
at $({\omega,\bf q})=0$ (which exhibits a dependence on angle in the $\{|\bq|,\omega\}$-plane) forces a profound modification of the standard analysis \cite{wilson,wilson_fisher,wilson_kogut}, as detailed in the next Section. 

\section{Renormalization Group procedure}
\subsection{Spherical harmonics decomposition}
In Euclidean space-time, the interaction in Eq.~(\ref{eq:Ueff}) decays as a power-law and depends on the angle relative to the time-like direction. This angular dependence exhibits a quadrupolar structure, parametrized by
\begin{align}
\sin^2\theta={v^2\bq^2\over \omega^2+v^2\bq^2}
\label{sin_theta_def}
\end{align}
where $v$ is the spin-wave velocity. To set the stage for a systematic RG analysis, we therefore perform a multipole-expansion of $U_{\mathrm{eff}}$ in terms of spherical harmonics in $D$ dimensions:
\begin{align}
U_{\mathrm{eff}}(\bq,\omega) = \sum_{n=0}^\infty u_nY_n(\theta)\, .
\label{eq:Ueffharmonics}
\end{align}
Recalling that we focus on $D=4-\epsilon$ where the interaction parameters $u_n$ near criticality are already of linear order in $\epsilon$, we use here the four-dimensional spherical harmonics $Y_n$ (see Appendix~\ref{app:AppA} for details). 
Note that although the spin-wave velocity $v$ flows under RG, the Gaussian part of the  action 
[the quadradic part of Eq.~(\ref{eq:Z})] is Lorentz invariant with respect to this velocity at every stage of the RG; hence 
it is natural to use the spherical harmonics expansion in terms of $\sin\theta$ as defined in Eq. (\ref{sin_theta_def}). 

When we perform the RG, we will find that higher even-order multipoles are generated at every step, which have the same naive scaling dimension.
The coefficients $u_n$ in Eq. (\ref{eq:Ueffharmonics}) will therefore be considered as an infinite set of running parameters. Their bare values are given by
\begin{align}
u_0&=\sqrt{2}\pi\lt(U_0+{c(c+2v)\over (c+v)^2}W\rt)
\end{align}
for $n=0$ and
\begin{align}
u_{n}&=-{2\sqrt{2}c\pi cv^2 (c-v)^{n-1}\over (c+v)^{n+2}}W
\end{align}
for $n\ge 1$.  They exhibit a systematic suppression by a velocity mismatch factor for increasing $n$:
\begin{align}
\frac{u_{n+1}}{u_{n}}= \frac{c-v}{c+v}\; .
\label{c_v_mismatch}
\end{align}
Note that for $c>v$, $W<0$ and $U_0>0$ imply that $u_n>0$ for all $n$. For $c<v$, $u_n$ of even and odd $n$ have alternating signs for $n\ge 1$; however, the leading phonon-induced coupling $u_1$ is always positive, as is $u_0$.


\subsection{Derivation of RG equations}
We now perform an RG scaling transformation following the $\epsilon$ expansion approach \cite{wilson,wilson_fisher,wilson_kogut}.  To this end we define a momentum shell 
\begin{equation}
{\Lambda\over b} < |\bk| < \Lambda
\end{equation}
which corresponds to the high wave-vector (short wave-length) fluctuations. We will denote momenta in this shell by $\bk^>$, and momenta $|\bk|<\Lambda/b$ by $\bk^<$. 
We then divide the integrals over the fields $\phi^\alpha(\bk,\omega)$ into fast and slow modes, $\phi^>$ and $\phi^<$, corresponding to modes with momenta $\bk\in\bk^>$ and $\bk\in\bk^<$ respectively.   Integration over the fast modes yields an effective action $S_\mathrm{eff}[\phi^<]$ for the slow modes, given by
\begin{eqnarray}
e^{-S_\mathrm{eff}[\phi^<]} &=& \int {\mathcal{D}} \phi^> e^{-S[\phi^>,\phi^<]}
\label{eq:Seff}
\end{eqnarray}
where
$\mathcal{D}\phi^>=\prod_{\omega}\prod_{\bk\in\bk^>}\prod_{\alpha=1}^N 
d\phi^\alpha(\bk,\omega)$.

\begin{figure}
\centering
\includegraphics[width=0.45\textwidth]{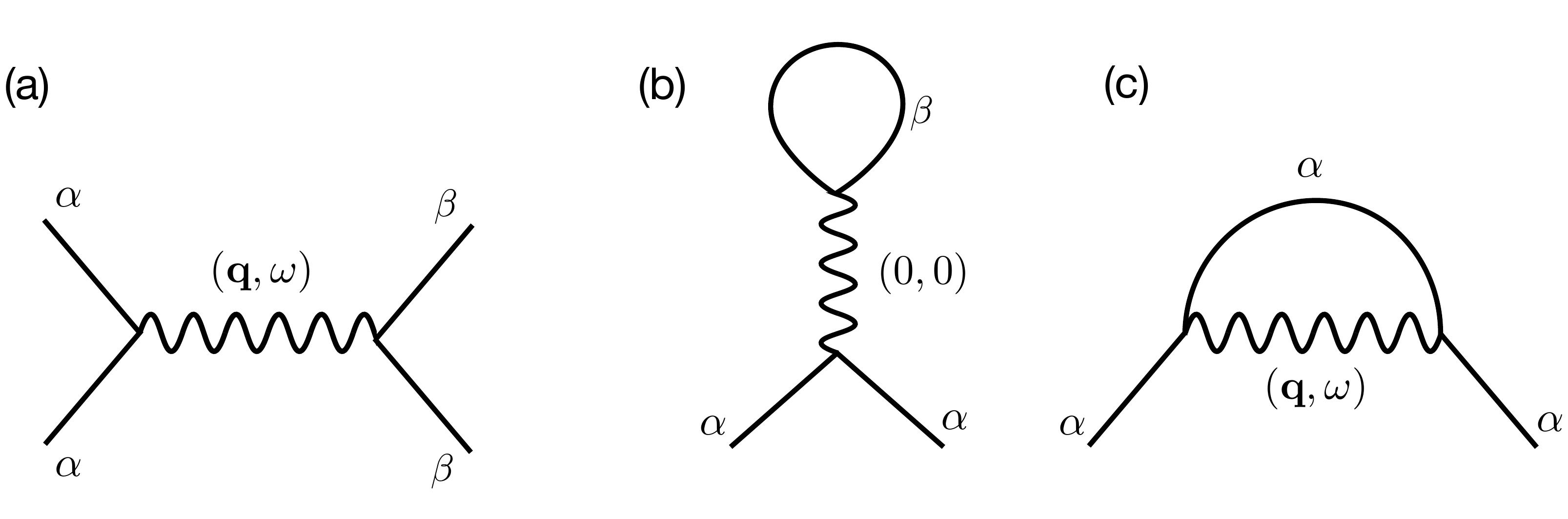}
\caption{(a): Feynman diagram corresponding to the interaction term in the action [Eq.~(\ref{eq:Z})]. Here the wiggly lines represent the effective interaction ${1\over N}U_{\mathrm{eff}}(\bq,\omega)$ [Eq. (\ref{eq:Ueff}) in the text]. The indices $\alpha, \beta$ correspond to different flavors of the $O(N)$-field propagators. (b)-(c): Feynman diagrams responsible for the renormalization of the terms quadratic in $\phi$ in the action [Eq.~(\ref{eq:Z})]. There are two possible diagrams, denoted $I_1$ (b) and $I_2$ (c). Note that $I_2$ depends on the transferred momentum and frequency ($\bq, \omega$).}
\label{fig:order1}
\end{figure}
The interaction $U_{\mathrm{eff}}$ in Eq.~(\ref{eq:Ueff}) gives rise to the Feynman rules shown in Fig.~\ref{fig:order1}(a).
Let us first consider the renormalization of the terms in the action that are quadratic in $\phi^<$.  These come from the Feynman diagrams shown in Fig.~\ref{fig:order1}(b) and \ref{fig:order1}(c), which we denote by $I_1$ and $I_2$. They are given by
\begin{align}
\label{eq:I1d}
I_1 &=  2\int_{\omega,\bk^>} {U_{\mathrm{eff}}({\bf 0},0)\over r+\omega^2+v^2\bk^2} \\
\label{eq:I2d}
I_2(\bk,\omega) &= 4{1\over N}\int_{\omega',\bk'^>} 
{U_{\mathrm{eff}}(\bk+\bk',\omega+\omega') \over r+\omega'^2+v^2\bk'^2}\,.
\end{align}
Note that $I_1$ [Eq.~(\ref{eq:I1d})] involves the interaction exactly at $(\bq=0,\omega=0)$.  In our case, this is ill-defined, since $U_{\mathrm{eff}}(\bk,\omega)$ depends on the direction in which the origin is approached.  Hence, we replace it by the spherical average over all directions, $\int 4\pi \sin^2(\theta) d\theta\, U_{\mathrm{eff}}(\theta)= 2\pi^2 u_0$.

To leading order in the $\epsilon$-expansion, we can set $d=3$ while evaluating $I_1$ and $I_2$ and we find
\begin{align}
\label{eq:I1}
I_1 &= {u_0\over 4\pi^2v^3}\lt(2v^2\Lambda^2-r\rt)\lt(1-{1\over b}\rt) \\
I_2(\bk,\omega) \!&=\! {1\over 2\pi^2v^3N}\lt[\lt(2v^2\Lambda^2-r\rt)u_0+2v^2\Lambda^2\sum_{n=1}^\infty u_n \rt. \no\\
&~~~~~~~~~~~~~~~~~~~~~~\lt. +{1\over 3}(3\omega^2-v^2k^2)u_1\rt]\lt(1-{1\over b}\rt) \no\\
&\equiv I_2(0,0)+A(3\omega^2-v^2k^2)
\label{eq:I2}\\
&\!\!\!\!\!\!\!\!\!\!\!\!\!\!\!\!\!\!\!
{\rm where}\quad A \equiv {1\over N}{u_1\over 6\pi^2v^3}\lt(1-{1\over b}\rt)\; .\no
\end{align}
Note that $I_2$ is momentum and frequency dependent. The expression in Eq.~(\ref{eq:I2}) was obtained by evaluating the integral in Eq.~(\ref{eq:I2d}) and then Taylor expanding it to order $k^2$ and $\omega^2$.  Higher order terms were discarded, since they are irrelevant in the RG sense.  Note that, despite the non-local nature of the phonon-mediated interaction, the shell integration of short-scale fluctuations leaves the kernel $I_2$ local, {\em i.e.} analytic in small $k$ and $\omega$, thus justifying the use of a Taylor series. In the leading order in $\epsilon$, the frequency and momentum dependence in Eq.~(\ref{eq:I2}) is only associated with the interaction $u_1$.  This term leads to renormalization of the spin-wave velocity $v$.

\begin{figure}
\centering
\includegraphics[width=0.49\textwidth]{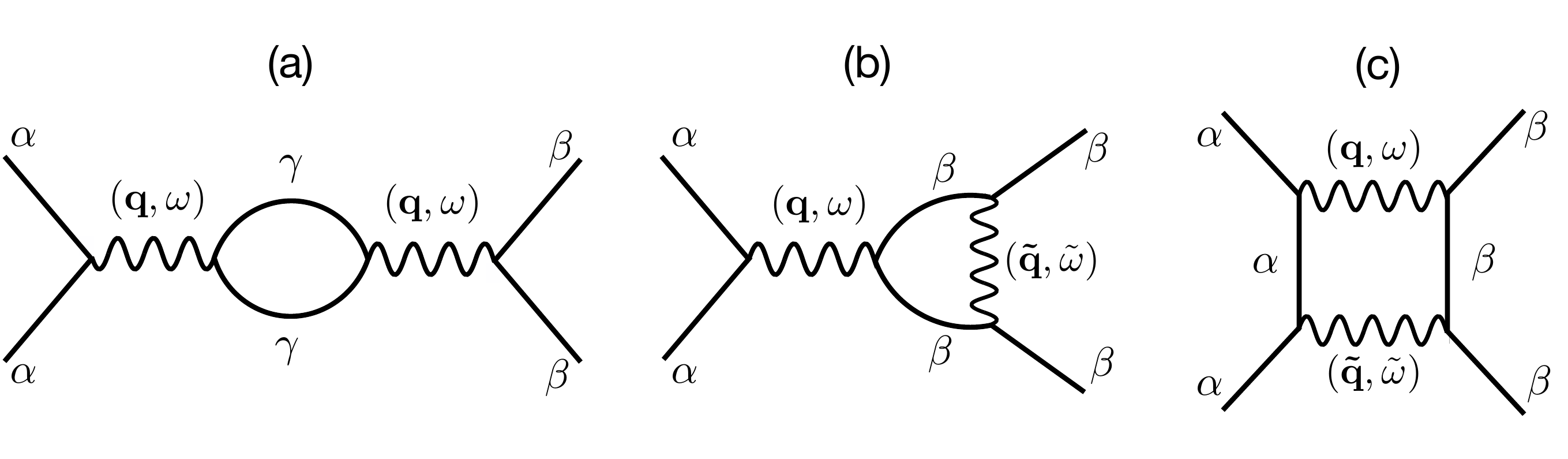}
\caption{Feynman diagrams arising from the renormalization of the quartic interaction terms in the action [Eq.~(\ref{eq:Z})]. Panels (a), (b) and (c) depict the three possible diagrams, denoted $D_1$, $D_2$ and $D_3$ respectively. While evaluating them, we set all external momenta to zero. 
}
\label{fig:2ndorder}
\end{figure}

Next, we will consider the renormalization of the quartic interaction.  This comes from three diagrams, shown in Fig.~\ref{fig:2ndorder}, which we denote by $D_1$, $D_2$, and $D_3$.
Once again accounting for the leading order in $\epsilon$, we calculate $D_{\{1,2,3\}}$ in three dimensions ($d=3$). Additionally, we set all external momenta to zero since any momentum dependence in $D_i$'s will be irrelevant, by power counting. 
Using these, we find [see Appendix~\ref{app:AppB} for details]
\begin{align}
\label{eq:D1}
D_1 =& {1\over \pi^2v^3N}\sum_{l,m=0}^\infty u_l u_m\sum_{n=|l-m|}^{l+m} Y_n(\theta) \lt(1-{1\over b}\rt) \\
\label{eq:D2}
D_2 =& {4\over \pi^2v^3N^2}u_0\sum_{n=0}^\infty u_n Y_n(\theta) \lt(1-{1\over b}\rt) \\
\label{eq:D3}
D_3 =& {4\over \pi^2v^3N^2}\sum_{n=0}^\infty u_n^2 \lt(1-{1\over b}\rt)\; .
\end{align}
Note that while evaluating the diagrams, we have used the fact that the product of two spherical harmonics is also a linear combination of spherical harmonics with the coefficients listed in Appendix~\ref{app:AppA}. Also note that while the result for $D_3$ contains only the zeroth harmonics $Y_0$ [Eq.~(\ref{eq:D3})], the diagrams $D_1$ and $D_2$ generate an infinite number of spherical harmonics [Eqs.~(\ref{eq:D1})-(\ref{eq:D2})]. However, since the higher harmonics are more rapidly oscillating, we work with up to the second harmonics $Y_2$, truncating for now the higher harmonics. Notably, we will show below that to leading order in the $\epsilon$-expansion, corrections arising from these higher harmonics do not alter the critical behavior.

Now incorporating the contributions from the one-loop diagrams, we can write the partition function as
\begin{widetext}
\begin{align}
\label{eq:Z2}
Z & =  \mathcal{N}\int {\mathcal{D}}\phi^<\
\exp\lt[ -{1\over 2} \int_{\omega,\bk^<} \lt\{r+\omega^2+v^2\bk^2+2\lt(I_1 +I_2(0,0)+A(3\omega^2-v^2k^2)\rt)\rt\} \!{\phi^<}^\alpha(\bk,\omega){\phi^<}^\alpha(-\bk,-\omega)\rt. \\
& \!\!\!\!\!\!\!\! \lt. -{1\over N}\!\int_{\omega,\bq^<}\!\int_{\omega_1,\bk_1^<}\!\int_{\omega_2,\bk_2^<}
\!\!\lt(\!U_{\mathrm{eff}}(\bq,\omega)\!-\!{N\over 2}\!\lt(D_1\!+\!D_2\!+\!D_3\rt)\!\!\rt)
\!{\phi^<}^\alpha({\rm \bf k_1},\omega_1){\phi^<}^\alpha({-{\rm \bf k_1}\!-\!{\rm \bf q}},-\omega_1\!-\!\omega){\phi^<}^\beta({{\rm \bf k_2}\!+\!{\rm \bf q}},\omega_2\!+\!\omega){\phi^<}^\beta(-{\rm \bf k_2},\!-\omega_2)\!\rt] \no
\end{align}
\end{widetext}
We then rescale ($\bk, \omega$) according to
\begin{align}
\bk &= b^{-1} \bk' \no\\ 
\omega &= b^{-1} \omega'
\label{eq:reskw}
\end{align}
which ensures that the upper bound of $\bk$ is restored back to $\Lambda$.
We subsequently rescale the fields $\phi^<$ according to
\begin{equation}
\phi=b^{(d+3)/2}(1+6A)^{-{1\over 2}}\phi'
\label{eq:resphi}
\end{equation} 
in order to keep the coefficient of $\omega^2$ in Eq.~(\ref{eq:Z2}) the same as in the original theory.
Note that the factor $A$ is proportional to $u_1$ [Eq. (\ref{eq:I2})], which governs the correction to the scaling dimension arising from the spin-phonon coupling. We will discuss this in detail in the Results section.

Using these rescalings in 
Eq.~(\ref{eq:Z2}), and setting $b=(1+dl)$, we obtain the RG equations to the leading order in $\epsilon$ as given in Appendix~\ref{app:AppC}:
\begin{widetext}
\begin{align}
\label{eq:RGr}
{dr\over dl} &= 2r+{1\over 2}\lt(1+{2\over N}\rt)(2v^2\Lambda^2-r)w_0 + {2v^2\Lambda^2-r\over N}w_1+{2v^2\Lambda^2\over N}w_2 \;,\\
\label{eq:RGv}
{dv\over dl} &=-{1\over N} {2vw_1\over 3} \;, \\
\label{eq:RGw0}
{dw_0\over dl} 
&= \epsilon w_0 -{1\over 2} \lt(1+{8\over N}\rt)w_0^2
-{1\over 2}\lt(1+{8\over 3N}\rt)w_1^2 -{1\over 2}\lt(1+{4\over N}\rt)w_2^2 \;, \\
\label{eq:RGw1}
{dw_1 \over dl} &= \epsilon w_1 -\lt(1+{2\over N}\rt)w_0w_1
-{1\over 2}\lt(1-{2\over 3N}\rt)w_1^2 -{1\over 2}w_2^2 -\lt(1-{1\over N}\rt)w_1w_2 \;, \\
\label{eq:RGw2}
{dw_2 \over dl} &= \epsilon w_2 -\lt(1+{2\over N}\rt)w_0w_2
-{1\over 2}\lt(1+{2\over 3N}\rt)w_1^2 -{1\over 2}w_2^2 -\lt(1-{1\over 3N}\rt)w_1w_2 \;.
\end{align}
\end{widetext}
Here, the new variables $w_n$ are directly related to the interaction parameters $u_n$ via a velocity-dependent rescaling:
 \begin{align}
w_n\equiv {u_n\over \pi^2v^3}
\end{align}
which simplifies the form of the final equations.

Note that the equations for the $w$'s do not depend on the other two parameters, $r$ and $v$, and hence close among themselves. Therefore, we can separately study the RG equations for the $w$'s, but keep in mind that at every RG stage, the solutions for the $w$'s impact the flow of $r$ and $v$. In the above RG equations, we truncated by eliminating the dependence on $w_3, w_4$ and so on [Eqs.~(\ref{eq:RGr})-(\ref{eq:RGw2})]. However, one can show that for all $n\ge 1$,
\begin{align}
{dw_n\over dl} &= \epsilon w_n-\lt(1+{2\over N}\rt)w_0w_n + \hdots
\label{eq:RGwn}
\end{align}
where $\hdots$ stands for terms that are bilinear in $w_l$ with $l\ge 1$. Later we will argue that the truncation will not affect our conclusions and we will discuss the relevance of Eq.~(\ref{eq:RGwn}). 

Recalling that the bare value of $w_1$ is positive (since $W<0$), Eq. (\ref{eq:RGv}) implies that the spin-wave velocity is initially renormalized downwards.  In practice, we find that $w_1$ does not change sign, hence this trend is maintained throughout the flow, and is only stopped in cases where $w_1$ flows to zero.

\section{Results}
When $W=0$, as is the case in the absence of phonons, all harmonics, except for $w_0$, vanish and the RG equations reduce to those of the standard O($N$) model. The underlying relativistic invariance then prevents $w_n$ with $n\ge 1$ from being generated in the RG flow. Therefore the Wilson-Fisher (WF) fixed point
\begin{align}
\label{eq:FPr}
v^*&=v_\mathrm{in} \,\,\, \mathrm{(initial\, value)}\\
r^* &= -\lt({N+2\over N+8}\rt)v_{\rm in}^2\Lambda^2\epsilon \\
w_0^* &=\lt({2N\over N+8}\rt)\epsilon \\
\label{eq:FPwn}
w^*_n &= 0 \text{~~for~~} n\ge 1
\end{align}
remains a fixed point of Eqs.~(\ref{eq:RGr})-(\ref{eq:RGw2}), although its stability can be affected by the phonon coupling. We note that the renormalized velocity, $v^*$, is not universal at the fixed point.

In order to determine whether the WF fixed point is stable to the addition of spin-phonon interactions, we linearize the RG equations around the fixed point, and obtain (considering up to $w_2$)
\begin{widetext}
\begin{equation}
{d\over dl}\!
\left[\!\!\begin{array}{ccccc}
\delta r\\ \delta v \\ \delta w_0 \\ \delta w_1 \\ \delta w_2
\end{array}\!\!\right] \!\!=\!\! 
\left[\!\begin{array}{ccccc}
2\!-\!{N+2\over N+8}\epsilon & ~4\!\lt({N+2\over N+8}\rt)\!v^*\!\Lambda^2\epsilon & \lt(1\!+\!{2\over N}\rt)\!\!\lt(1\!+\!{N+2\over 2(N+8)}\epsilon\rt)\!{v^*}^2\!\Lambda^2 & ~{2\over N}\!\!\lt(\!1\!+\!{N+2\over 2(N+8)}\epsilon\rt)\!{v^*}^2\!\Lambda^2 & {2{v^*}^2\Lambda^2\over N} \\
0 & 0 & 0 & -{2{v^*}\over 3N} & 0 \\
0 & 0 & -\epsilon & 0 & 0 \\
0 & 0 & 0 & {4-N\over N+8}\epsilon & 0 \\
0 & 0 & 0 & 0 & {4-N\over N+8}\epsilon
\end{array}\!\!\right]
\!\!\!\!\left[\!\!\begin{array}{ccccc}
\delta r \\ \delta v \\ \delta w_0 \\ \delta w_1 \\ \delta w_2 
\end{array}\!\!\right]
\label{eq:RGlinear}
\end{equation}
\end{widetext}
Since the matrix is upper-triangular, the eigenvalues are given by the diagonal elements.  In fact, based on Eq. (\ref{eq:RGwn}) one can see that the upper-triangular structure of the matrix is preserved when all $w_n$s are kept and no truncation is done.  Then, the diagonal element corresponding to $\delta w_n$ takes the value
\begin{align}
\frac{4-N}{N+8}\epsilon
\end{align} 
for all $n\ge 1$.  This demonstrates that $N=4$ is special, and that the WF fixed point is stable for $N>4$ and unstable for $N<4$.  We will next consider the cases $N>4$ and $N<4$ separately.

\subsection{Case $N>4$}
For $N>4$, the WF fixed point is stable.
Since $\delta w_n$ decays exponentially with $l$ for all $n\ge 1$, $w_n$ is marginally irrelevant and the WF fixed point is robust against small spin-phonon coupling.  


We furthermore argue that the critical exponents are identical to the standard O($N$) model.  To see this, note that the matrix in Eq. (\ref{eq:RGlinear}) is upper triangular, and therefore its eigenvalues are given by its diagonals.  In particular, the scaling dimension of $r$, $y_r=2-{N+2\over N+8}\epsilon$, yields the known value for the correlation length exponent $\nu$.  Then, the remaining exponents ($\beta$, $\gamma$, {\em etc.}), can be deduced from the scaling of the field $\phi$ [Eq. (\ref{eq:resphi})].  This differs from the WF by a factor dependent on $A$ which in turn is proportional to $w_1$ [Eq. (\ref{eq:I2})].  Since $w_1^*=0$, this reduces to the standard scaling of the O($N$) model.

\subsection{Case $N<4$}
For $N<4$, $w_n$ is relevant and hence starting from a small initial positive value ${w_n}_{\rm in}$ it grows rapidly for all $n\ge 1$. Note that all $w_n$ have a similar behavior, however in the following discussion we will mainly focus on $w_1$ since it corresponds to the leading spin-phonon coupling. In order to search for a fixed point that describes the new critical point, we solve the RG equations and find that one of the fixed points is the WF one as described in Eqs.~(\ref{eq:FPr})-(\ref{eq:FPwn}). We find that, regardless of the initial values of the parameters, the RG equations become unstable, indicating a first-order transition. The source of this instability is that, even for arbitrarily small initial value ${w_1}_{\rm in}$, $w_1$ eventually grows rapidly 
and drives $w_0$ to become \textit{negative} -- see the last two terms in Eq.~(\ref{eq:RGw0}). Once $w_0$ is negative, the $\phi^4$ theory collapses. 

\begin{figure}
\includegraphics[width=0.7\columnwidth]{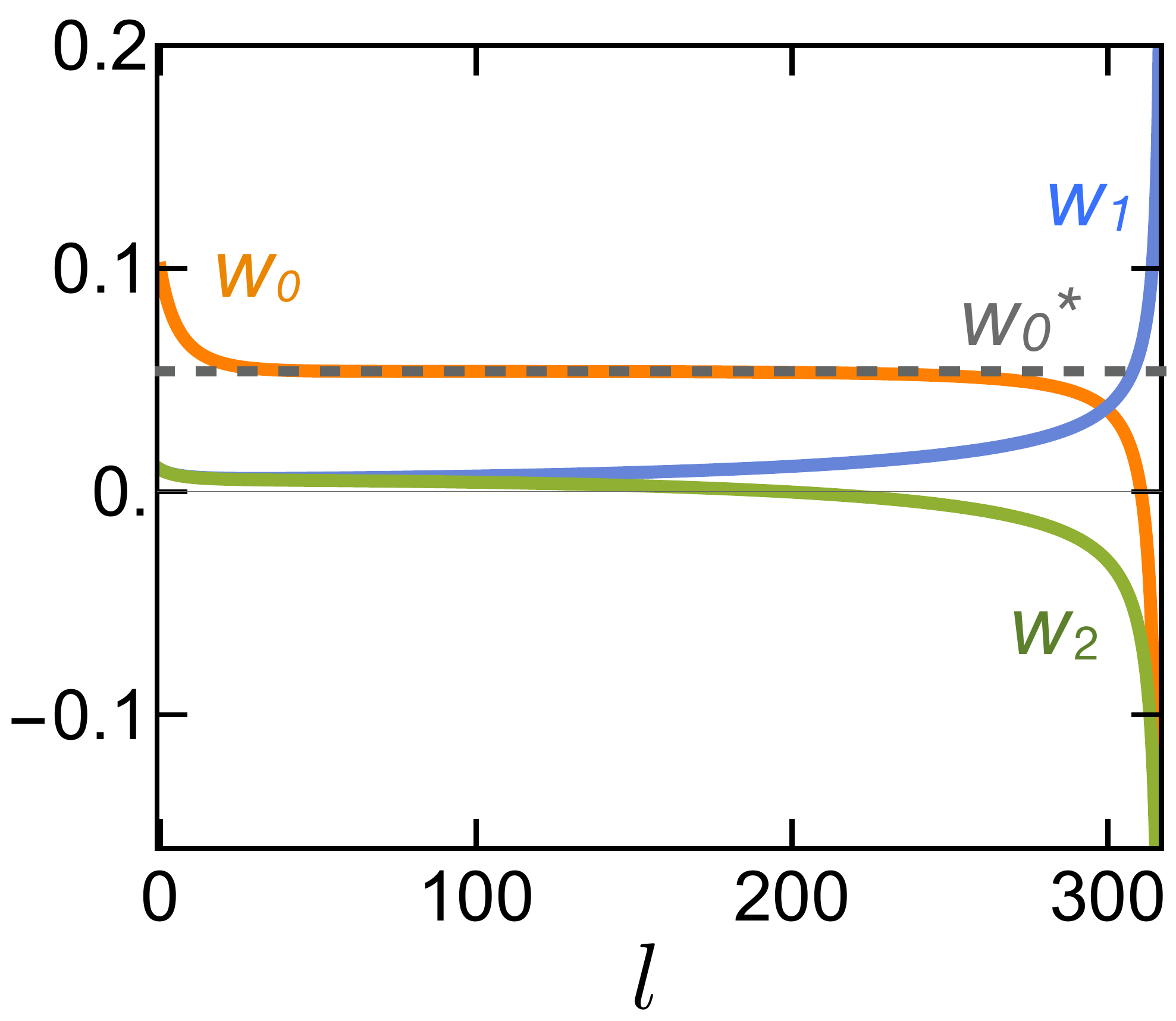}
\caption{ The flow of $w_0$, $w_1$ and $w_2$ are shown as a function of RG time $l$, along with the Wilson-Fisher value of $w_0$ {\em i.e.} $w_0^*$ for $N=3$. Starting from an initial value of ${w_0}_{\rm in}=0.1$, it first decays exponentially and reaches a constant value. $w_1$ on the other hand starts from a small positive value (${w_1}_{\rm in}=0.01$), then grows fast and eventually pulls $w_0$ to a negative value. $w_2$ also starts from the same initial value ${w_2}_{\rm in}=0.01$, but it turns to a negative value and diverges near the instability. $w_0$ spends a long time at its Wilson-Fisher value showing a plateau. We define a characteristic scale $l^*$ where the above described breakdown occurs (see text).}
\label{Fig:flow}
\end{figure}

Nevertheless, for ${w_1}_{\rm in}$ sufficiently close to $w_1^*=0$, the WF fixed point still plays an important role.  Then, $w_1$ remains small for a long RG time $l$, during which its effect on other parameters can be neglected.  As a result, $w_0$ flows towards its WF value  $w_0^*$, and stays there for a long time, as shown in Fig.~\ref{Fig:flow}. Eventually, when $w_1$ grows sufficiently, it pulls $w_0$ to negative values, at which point the RG equations become unstable.

The fact that the system spends a long time near the fixed point implies that the transition is weakly first-order.  This follows since the value $l^*$ of the RG parameter $l$ at the instability is large, and therefore the characteristic scale $\xi^*\sim e^{l^*}$ at the instability is also large. The picture that then emerges is that, as one approaches the phase transition between ordered and disordered phases, the correlation length grows following the standard WF exponent. However, close enough to the transition, the divergence in the correlation length is cut off by the scale $\xi^*$.
Thus, $\xi^*$ provides the characteristic correlation length at the first-order transition.

Here we provide an estimate for the correlation length $\xi^*$. We note that the instability occurs when 
$w_1$ grows and becomes roughly equal to $w_0^*$. We assume that starting from a small initial value ${w_1}_{\rm in}$, $w_1$ grows exponentially (as in its linear order)
\begin{equation}
w_1 = {w_1}_{\rm in}e^{{(4-N)\over(N+8)}\epsilon l}
\end{equation}
until the system reaches this instability. Using this, we obtain the estimated correlation length
\begin{align}
\xi^* &= ae^{l^*} \approx a\lt[{2N\over N+8}{\epsilon\over {w_1}_{\rm in}}\rt]^{{(N+8)\over(4-N)}{1\over\epsilon}}
\label{eq:xi}
\end{align}
where $a$ is the lattice constant. We have checked that for small $\epsilon$ and small ${w_1}_{\rm in}$, this analytic estimate of the correlation length is consistent with the length-scale obtained from numerics, at which the value of $v$ sharply falls to zero, and $w_0$ and $w_1$ diverge.
The weakly first-order transition in presence of spin-phonon coupling can be characterized by this length-scale $\xi^*$. For system sizes up to this length-scale the system exhibits correlations and scaling consistent with the Wilson-Fisher second-order transition.  Beyond that scale, a weakly first-order transition is manifested.  From Eq.~(\ref{eq:xi}), we see that $\xi^*$ grows exponentially when either $\epsilon$ or the coupling to phonons become smaller. Note that the picture remains qualitatively same when including higher harmonics in the RG equations.

Although the quantum critical point is strictly defined at zero temperature, in any physical realization there is a finite temperature which is related to the finite size in the Euclidean time direction. Therefore, there is a temperature scale $T^*$ associated with this length-scale, given by $T^*\sim {\hbar v\over k_B\xi^*}$.  Above this temperature, the quantum RG flow is cut-off by the finite size in the time direction while $w_1$ is still small and $w_0$ is still positive. Then, the system is described by a thermal transition in $d=3-\epsilon$ space dimensions, which for small enough $w_1$ may result in a second-order transition \cite{aharony}. $T^*$ is therefore the temperature of the tricritical point separating second-order and weakly first-order transitions [see Fig.~\ref{Fig:phasediag}(b)].


Finally, we consider the situation where the number of flavors is exactly 4 ($N=4$). In this marginal case, we numerically find that the transition is always weakly first-order, no matter how weak ${w_1}_{\rm in}$ is. This is consistent with the observation that, even for $N>4$, the transition becomes first-order whenever ${w_1}_{\rm in}$ exceeds some $N$-dependent threshold value ${w_1}_{\rm thr}(N)$.  We find that, as $N$ approaches 4 from above, ${w_1}_{\rm thr}(N)$ approaches zero, indicating that at $N=4$ the second-order transition is first-order even for infinitesimal phonon coupling.

\subsection{Other fixed points}
We now consider the other possible fixed points, which are different from the Wilson-Fisher one. We find that additional fixed points do exist, but they are unstable and therefore do not play an important role in the RG, as explained below.

We first consider two RG equations Eqs.~(\ref{eq:RGw0})-(\ref{eq:RGw1}) in the absence of $w_2$. We can analytically solve these two equations to obtain the fixed points:
\begin{align}
\label{eq:FPw0pm}
{w_0^*}^{(\pm)} &\!\!=\! {\!N\!\!\lt(\!196\!+\!\!156N\!\!+\!45N^2\!\pm\! \sqrt{5}(2\!-\!3N)\!\sqrt{9N^2\!\!-\!\!12N\!\!-\!\!76}\rt)\!\epsilon \over 416+436N+300N^2+45N^3} \\
\label{eq:FPw1pm}
{w_1^*}^{(\pm)} &\!\!=\! 2{\epsilon - \lt(1+{2\over N}\rt){w_0^*}^{(\pm)} \over \lt(1-{2\over 3N}\rt)} \;.
\end{align}
Note that these two fixed points are real when the number of flavors is larger than a critical number obtained by solving the equation $9\tilde N_c^2-12\tilde N_c-76=0$, i.e. $N>\tilde N_c\approx 3.64$.
We numerically find that the fixed point FP$^{(+)}$ is unstable in both directions for all $N>\tilde N_c$. On the other hand, the fixed point FP$^{(-)}$ starts from a positive value of $w_1^*$ at $N=\tilde N_c$, coincides with the WF fixed point at $N=4$ (with $w_1^*=0$), and turns negative thereafter. We also notice that this fixed point is stable in both directions in the $\{w_0,w_1\}$ plane in the range $\tilde N_c<N<4$.

These results relate to the truncated model with two interaction parameters ($w_0$ and $w_1$), but adding extra parameters could give rise to unstable directions.  To test this, we add the higher order harmonic $w_2$, and numerically find the new fixed points [to  Eqs.~(\ref{eq:RGw0})-(\ref{eq:RGw2})] and study their stability. We still find two fixed points.  However, with the addition of the third parameter $w_2$, the value of $\tilde N_c$ decreases to 2.48, and an unstable direction appears in the FP$^{(-)}$ fixed point.  We find no reason to expect that the conclusion might change by adding higher-order harmonics, as this increases the potential for more unstable directions. We therefore conclude that there is no stable fixed point in the system apart from the WF fixed point for $N>4$.


\section{Summary and outlook}
In summary, using renormalization group analysis in $4-\epsilon$ dimensions, we have studied the quantum phase transition in presence of acoustic phonons. We have shown that when the number of flavors of the underlying O($N$) model is larger than a critical number $N_c=4$, the transition remains a standard second-order one. On the other hand when $N<N_c$, the transition becomes weakly first-order one, characterized by a large length-scale $\xi^*$ or, equivalently, by a small temperature scale below which the transition changes from second-order to first-order.  We are currently in the process of verifying these analytical predictions numerically, using Monte Carlo simulations of O($N$) models coupled to phonons.

Throughout our analysis, we ignored the feedback of the O($N$) field on the phonons.  This is justified provided that the phonons are stiff enough and the coupling to phonons is not too strong.  However, it would be interesting to understand situations where these assumptions may break down, {\em e.g.} for systems near structural transitions where the phonons are softened and susceptible to non-linear corrections arising from coupling to the O($N$) fluctuations.
Indeed, when the coupling to the phonons is strong enough, new phases involving structural reorganization of the lattice can occur.  For example, in Ref.~[\onlinecite{santiago}] this was demonstrated for an Ising model strongly coupled to optical phonons. It would be interesting to understand how this feedback can affect critical properties.

\begin{acknowledgements}
The authors acknowledge helpful discussions with Amnon Aharony, Ehud Altman, Premala Chandra, Snir Gazit and Arun Paramekanti. DP thanks the Israel Science Foundation for financial support (grant 1803/18). ES thanks the Aspen Center for Physics (NSF Grant No. 1066293)
for its hospitality, and financial support by the
US-Israel Binational Science Foundation through
awards No. 2016130 and 2018726, and by the Israel Science Foundation (ISF) Grant No. 993/19.
\end{acknowledgements}
\appendix
\section{Spherical harmonics in (3+1) dimensions}
\label{app:AppA}
We introduce the even-order spherical harmonics in 4 dimensions ($D=d+1=4$), the first few of which are:
\begin{align}
Y_0(\theta) &= {1}\; , \no\\
Y_1(\theta) &= -4 \sin^2\theta+3\; , \no\\
Y_2(\theta) &= 16\sin^4\theta-20\sin^2\theta+5\; , \no\\
Y_3(\theta) &= -64\sin^6\theta+112\sin^4\theta-56\sin^2\theta+7\; , \no\\
&\vdots 
\end{align}
where $\theta$ is the angle relative to the vertical $\omega$ axis in the $(v|q|,\omega)$ plane,
\begin{align}
\sin^2\theta &= {v^2\bq^2\over \omega^2+v^2\bq^2} \; .
\end{align}
These functions are found constructively: we choose $Y_0(\theta)=1$.  Then, for $n>1$, $Y_n$ is an even polynomial in $\sin\theta$ of order $2n$, whose coefficients are fixed by orthonormalizing it with all $Y_l$ of lower order,
\begin{align}
\lt(Y_l,Y_n\rt)=\delta_{ln}\;.
\end{align}
Here, the inner product of two real functions $f(\theta)$ and $g(\theta)$ is defined by,
\begin{align}
\lt(f,g\rt)\equiv  {2\over\pi}\int_0^\pi \sin^2\theta\, d\theta f(\theta)\,g(\theta)\;,
\end{align}
Note that our integration measure
\begin{align}
d\Omega=\frac{2}{\pi}\sin^2\theta\, d\theta    
\end{align}
differs from the standard integration measure for angular integrals in four-dimensions,  ${4\pi}\sin^2\theta$, by an overall factor of $2\pi^2$.  This choice is convenient as it simplifies many of the expressions that follow.

The functions $Y_n$ depend only on $\theta$, the angle from the time-like axis.  In particular, they are SO(3)$_{\rm space}$ invariant, {\em i.e.} they do not depend on the orientation in the three-dimensional, space-like, directions.  Thus, these functions are the four-dimensional analogues of the azimutally symmetric ($m=0$), even order, spherical harmonics in 3D, $Y_{\ell m}$, with $\ell=2n$ and $m=0$.  There are also odd harmonics, involving odd powers of $\sin\theta$, and harmonics that are not SO(3)$_{\rm space}$-invariant.  However, since the bare action is even under parity, and since it is SO$_{\rm space}$ invariant, these harmonics are not generated in the RG.

Using trigonometric identities, the functions above can be rewritten in a simple form,
\begin{align}
Y_0(\theta) &= {1}\; , \no\\
Y_1(\theta) &= 1+2\cos(2\theta)\; , \no\\
Y_2(\theta) &= 1+2\cos(2\theta)+2\cos(4\theta)\; , \no\\
&\vdots \no\\
Y_n(\theta) &= 1+2\cos(2\theta)+...+2\cos(2n\theta)\; 
\end{align}
In this form, and writing the integration measure as $d\Omega=\frac{1}{\pi}\lt(1-\cos(2\theta)\rt)$, it is straightforward to show that the functions are an orthonormal set.

These functions can be used to expand any even, SO(3)$_{\rm space}$-invariant, function.  Note that the product of two spherical harmonics $Y_l(\theta)Y_m(\theta)$ is itself another even, SO(3)$_{\rm space}$-symmetric function.  Therefore, it can be expanded in terms of $Y_n(\theta)$s,
\begin{align}
Y_l(\theta)Y_m(\theta)=\sum_{n=0}^\infty a_{n;lm}Y_n(\theta)\,.
\end{align}
One can show that the coefficients $a_{n;lm}$ are given by
\begin{align}
a_{n;lm} = \begin{cases}
1 &\, \text{for~~ $|l-m|\le n\le l+m$}\\
0 &\, \text{otherwise}
\end{cases}
\end{align}

We now use these results to derive a number of relations which will be useful when computing Feynman diagrams.  Expanding the interaction as a linear combination of the spherical harmonics,
\begin{align}
U_{\mathrm{eff}}(\bq,\omega) &= \sum_{n=0}^\infty u_nY_n(\theta) \; .
\end{align}
we obtain
\begin{align}
\label{eq:omint1}
\int d\Omega \, U_{\mathrm{eff}}(\bq,\omega) &= u_0 \; ,\\
\label{eq:omint2}
\int d\Omega \, \lt(U_{\mathrm{eff}}\rt(\bq,\omega))^2 &= \sum_{n=0}^\infty u_n^2 \; ,
\end{align}
and
\begin{align}
\lt(U_{\mathrm{eff}}(\bq,\omega)\rt)^2=&\sum_{l,m=0}^\infty u_l u_m Y_l(\theta)Y_m(\theta) \no\\ =&\sum_{l,m=0}^\infty u_l u_m\sum_{n=|l-m|}^{l+m} Y_n(\theta) \; .
\label{eq:omint3}
\end{align}
We will use these results while evaluating the Feynman diagrams $D_i$'s.

\section{Evaluation of the Feynman diagrams arising from the renormalization of the interaction}
\label{app:AppB}
Renormalization of the quartic interaction term in the action [Eq.~(\ref{eq:Z})] comes from the diagrams $D_1$, $D_2$ and $D_3$ shown in Fig.~\ref{fig:2ndorder}. Here we provide the integrals involved in those diagrams:
\begin{align}
\label{eq:intD1}
D_1(\bk,\omega) &= {8\over N}\left(U_{\mathrm{eff}}(\bk,\omega)\right)^2\int_{\omega',\bk'^>} {1\over {(r+\omega'^2+v^2\bk'^2)^2}} \\
\label{eq:intD2}
D_2(\bk,\omega) &= {32\over N^2}U_{\mathrm{eff}}(\bk,\omega)\int_{\omega',\bk'^>} {U_{\mathrm{eff}}(\bk',\omega')\over {(r+\omega'^2+v^2\bk'^2)^2}} \\
\label{eq:intD3}
D_3 &= {32\over N^2}\int_{\omega',\bk'^>} {U_{\mathrm{eff}}(\bk',\omega')^2\over {(r+\omega'^2+v^2\bk'^2)^2}} \; .
\end{align}
In order to evaluate these integrals we set $r=0$, since it always contributes corrections $\sim\mathcal{O}\lt({r\over \Lambda^2}\rt)\ll 1$. 

\begin{figure}[h!]
	\includegraphics[width=0.8\columnwidth]{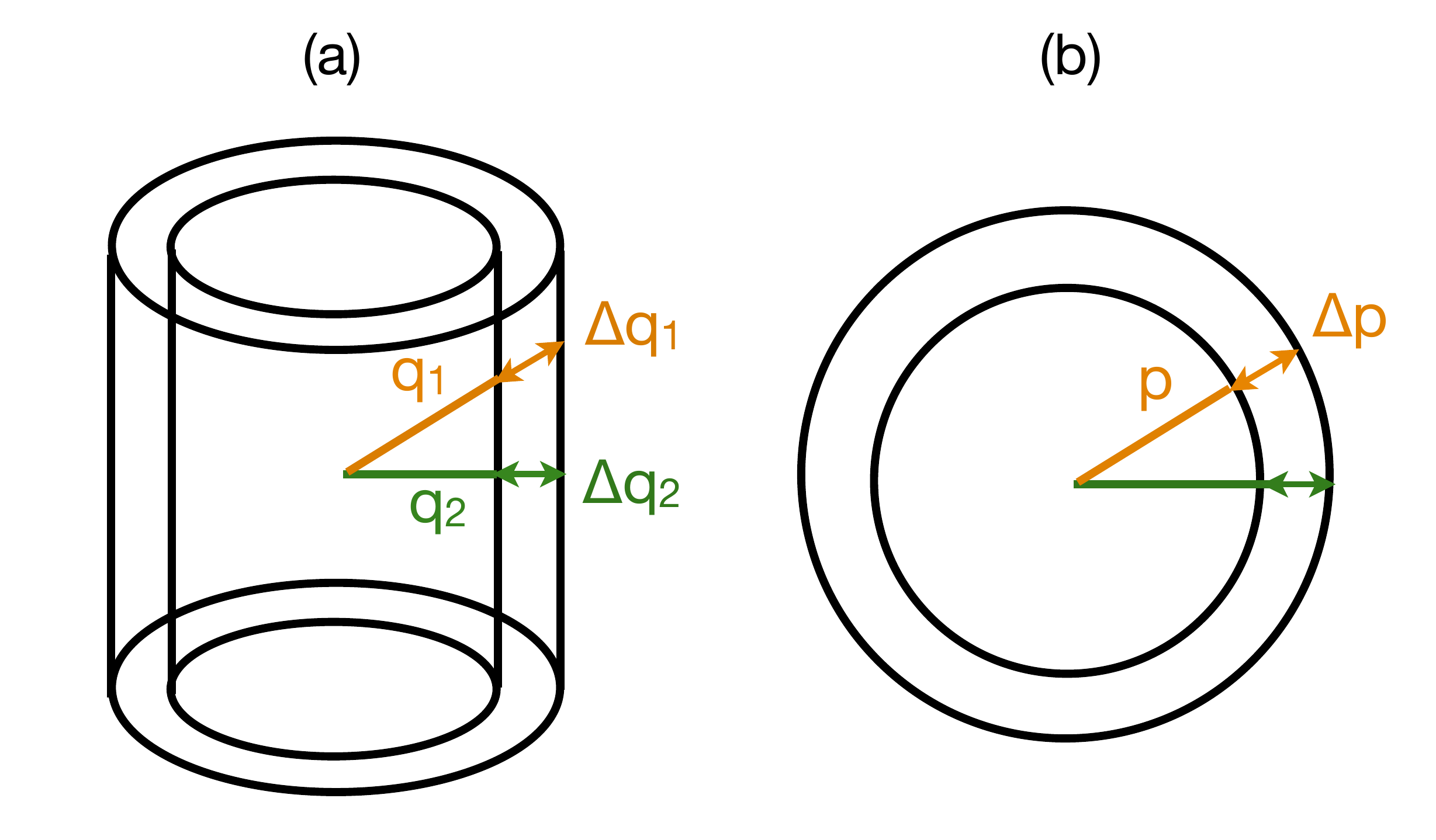}
	\caption{(a) Cylindrical shell.  We use an angle-dependent rescaling transformation, $p^\mu(\Omega)=g(\Omega)q^\mu$, and choose the rescaling function $g(\Omega)$ to deform the cylindrical shell to a spherical shell, shown in (b).  Since $\frac{\Delta q_1}{q_1}=\frac{\Delta q_2}{q_2}$, the resulting spherical shell has constant thickness.}
	\label{Fig:Omega}
\end{figure}
We now introduce the Euclidean 4-vector
$q^\mu\equiv (\omega',v\bk')$, and note that $U_{\mathrm{eff}}(q^\mu)$ depends on the direction $\Omega$ of $q^\mu$, but not on its magnitude, $q$.  Then, one can see that
all of the above integrals take the following form:
\begin{align}
\int_{\rm cylinder} {d^4q\, f(\Omega)\over q^4} =2\pi^2 \int_{\rm cylinder} d\Omega f(\Omega)\int{dq\ q^3 \over q^4}
\label{eq:cylInt}
\end{align}
integrated over the cylindrical shell in Fig.~\ref{Fig:Omega}(a).  In particular, the domain of the $q$ integral depends on $\Omega$.

We next introduce a direction-dependent scaling factor, $g(\Omega)$, and a rescaled variable $p^\mu=g(\Omega) q^\mu$, such that the cylindrical shell in $q^\mu$, Fig.~\ref{Fig:Omega}(a), is deformed to a spherical shell in $p^\mu$, Fig.~\ref{Fig:Omega}(b).  Note that the thickness of the cylindrical shell, $\Delta q$, is proportional to the distance of the shell to the origin, $q$, such that if two different directions $\Omega_1$ and $\Omega_2$ are compared
\begin{align}
    \frac{\Delta q_1}{q_1}=\frac{\Delta q_2}{q_2}\, .
\end{align}
This implies that the spherical shell in Fig.~\ref{Fig:Omega}(b) has constant thickness. Furthermore, since $\frac{dp}{p}=\frac{g(\Omega)dq}{g(\Omega)q}=\frac{dq}{q}$, the integral
in Eq. (\ref{eq:cylInt}) becomes
\begin{align}
\int_{\rm spher. shell}  d\Omega f(\Omega)\int{dp \over p}\, .
\label{eq:spherInt}
\end{align}
In this form, the domain of integration is spherically symmetric.  In particular, the domain of the $p$ integral is independent of $\Omega$, and can be evaluated directly
\begin{align}
\int_{\Lambda/b}^\Lambda \frac{dp}{p}=\log b\approx 1-\frac{1}{b}
\end{align}
where the final result is correct to linear order in $b-1$ (which equals $dl$).  Now, to compute the angular integral $\int d\Omega\,f(\Omega)$, all the results derived in Appendix \ref{app:AppA},  Eqs.~(\ref{eq:omint1})-(\ref{eq:omint3}), can be used.  Combining these results, we end up with the final answers for $D_{1,2,3}$ given in Eqs.~(\ref{eq:D1})-(\ref{eq:D3}) of the main text.

\section{Derivation of the RG equations}
\label{app:AppC}
In this Appendix, we provide the intermediate steps to obtain the RG equations given in the main text [Eqs.~(\ref{eq:RGr})-(\ref{eq:RGw2})]. We apply the rescaling of $(\bk,\omega)$ [Eq.~(\ref{eq:reskw})] and the scalar field [Eq.~(\ref{eq:resphi})] in Eq.~(\ref{eq:Z2}). Then, comparing the coefficients of different parameters with those in Eq.~(\ref{eq:Z}), we obtain the following expressions for the renormalized parameters:
\begin{widetext}
\begin{align}
r' &= {b^2\over 1+6A}\lt(r + 2I_1 + 2I_2(0,0)\rt) \;, \\
v'^2 &= \lt(1-2A\over 1+6A\rt)v^2 \;, \\
u_0' &= {b^{3-d}\over (1+6A)^2} \lt[u_0-{u_1\over v}{dv\over dl}dl-{1\over 2\pi^2v^3} \lt\{\lt(1+{8\over N}\rt)u_0^2+\lt(1+{4\over N}\rt)(u_1^2+u_2^2)\rt\}\lt(1-{1\over b}\rt)\rt] \;, \\
u_1' &= {b^{3-d}\over (1+6A)^2} \lt[u_1-{{u_1+3u_2}\over 2v}{dv\over dl}dl-{1\over 2\pi^2v^3} 
\lt\{\lt(2+{4\over N}\rt)u_0u_1+(u_1+u_2)^2\rt\}\lt(1-{1\over b}\rt) \rt] \;, \\
u_2' &= {b^{3-d}\over (1+6A)^2} \lt[u_2+{{u_1-u_2}\over 2v}{dv\over dl}dl-{1\over 2\pi^2v^3} 
\lt\{\lt(2+{4\over N}\rt)u_0u_2+(u_1+u_2)^2\rt\}\lt(1-{1\over b}\rt) \rt] \;,
\end{align}
Note that the above renormalized parameters $u_n'$ also include terms proportional to ${dv\over dl}$.  These arise due to the evolution of the harmonics $Y_n$ as the spin-wave velocity changes with the RG flow, which is of the form
\begin{align}
\lt[\sum_{n=0}^\infty u_nY_n\rt]_{l+dl} = \lt[\sum_{n=0}^\infty u_nY_n\rt]_l +\lt[\sum_{n=0}^\infty u_n{dY_n\over dv}\rt]_l{dv\over dl}dl
\end{align}
and we use the identities ${dY_0(\theta)\over dv}=0$ and ${dY_n(\theta)\over dv}={1\over 2v}\lt[-(n+1)Y_{n-1}(\theta)-Y_n(\theta)+nY_{n+1}(\theta)\rt]$ (valid for $n\ge 1$) to evaluate the second term. Finally, setting $b=(1+dl)$ and using the functional forms of $A,I_1$ and $I_2(0,0)$ as given in Eqs.~(\ref{eq:I1})-(\ref{eq:I2}), lead to the following RG equations:
\begin{align}
{dr\over dl} &= 2r+\lt(1+{2\over N}\rt){(2v^2\Lambda^2-r)u_0\over2\pi^2v^3} + {(2v^2\Lambda^2-r)u_1\over N\pi^2v^3}+{2\Lambda^2\over N\pi^2v}u_2~, \\
{dv\over dl} &= -{1\over N} {2u_1\over 3\pi^2}{1\over v^2} \;, \\
{du_0\over dl} &= \epsilon u_0 -{1\over 2\pi^2v^3} \lt(1+{8\over N}\rt)u_0^2
-{1\over N} {2\over \pi^2v^3}u_0u_1-{1\over 2\pi^2v^3}\lt(1+{8\over 3N}\rt)u_1^2 
-{1\over 2\pi^2v^3}\lt(1+{4\over N}\rt)u_2^2 \;, \\
{du_1 \over dl} &= \epsilon u_1  -{1\over \pi^2v^3}\lt(1+{2\over N}\rt)u_0u_1
-{1\over 2\pi^2v^3}\lt(1+{10\over 3N}\rt)u_1^2
-{1\over 2\pi^2v^3}u_2^2 -{1\over \pi^2v^3}\lt(1-{1\over N}\rt)u_1u_2 \;, \\
{du_2 \over dl} &= \epsilon u_2  -{1\over \pi^2v^3}\lt(1+{2\over N}\rt)u_0u_2
-{1\over 2\pi^2v^3}\lt(1+{2\over 3N}\rt)u_1^2
-{1\over 2\pi^2v^3}u_2^2 -{1\over \pi^2v^3}\lt(1+{5\over 3N}\rt)u_1u_2
\end{align}
Now we define $w_n={u_n\over \pi^2v^3}$, which implies ${dw_n\over dl}={1\over \pi^2}\lt[{1\over N}{2u_1u_n\over v^6}+{1\over v^3}{du_n\over dl}\rt]$, and using these the above equations simplify to Eqs.~(\ref{eq:RGr})-(\ref{eq:RGw2}) in the main text.
\end{widetext}


\end{document}